# Probable Gravitational Microlensing Towards the Galactic Bulge

C. Alcock[*,†], R.A. Allsman[*], T.S. Axelrod[*],
D.P. Bennett[*,†], S. Chan[‡], K.H. Cook[*,†], K.C. Freeman[‡],
K. Griest[†,∥], S.L. Marshall[†,♭], S. Perlmutter[†],
B.A. Peterson[‡], M.R. Pratt[†,♭], P.J. Quinn[‡], A.W. Rodgers[‡],
C.W. Stubbs[†,♭], W. Sutherland[♠]
(The MACHO Collaboration)

[*] Lawrence Livermore National Laboratory, Livermore, CA 94550

[†] Center for Particle Astrophysics, University of California, Berkeley, CA 94720

[‡] Mt. Stromlo and Siding Spring Observatories,
Australian National University, Weston, ACT 2611, Australia

[∥] Department of Physics, University of California, San Diego, CA 92039

[♭] Department of Physics, University of California, Santa Barbara, CA 93106

[♠] Department of Physics, University of Oxford, Oxford OX1 3RH, U.K.

## Abstract

The MACHO project carries out regular photometric monitoring of millions of stars in the Magellanic Clouds and Galactic Bulge, to search for very rare gravitational microlensing events due to compact objects in the galactic halo and disk. A preliminary analysis of one field in the Galactic Bulge, containing ∼ 430,000 stars observed for 190 days, reveals four stars which show clear evidence for brightenings which are time-symmetric, achromatic in our two passbands, and have shapes consistent with gravitational microlensing. This is significantly higher than the ∼ 1 event expected from microlensing by known stars in the disk. If all four events are due to microlensing, a 95% confidence lower limit on the optical depth towards our bulge field is $1.3 \times 10^{-6}$, and a "best fit" value is $\tau \approx 1.6 \times 10^{-6}/\epsilon$, where $\epsilon$ is the detection efficiency of the experiment, and $\epsilon < 0.4$. If the true optical depth is close to the "best fit" value, possible explanations include a "maximal" disk which accounts for most of the galactic circular velocity at the solar radius, a halo which is centrally concentrated, or bulge-bulge microlensing.

# 1. Introduction

Gravitational microlensing is a very powerful probe of the mass distribution around our Galaxy, since it is sensitive to any population of compact objects independent of their emission of electromagnetic radiation. Paczyński (1986) proposed that gravitational microlensing of stars in the Magellanic Clouds could provide a test for dark matter in the form of Massive Compact Halo Objects (MACHOs), via the transient brightening which results as a MACHO passes near the line of sight to a distant star. These events are expected to be extremely rare, with optical depth $\tau \lesssim 10^{-6}$, so a very large number of stars ($\sim 10^6$) must be monitored for many months to provide statistically significant results. Variable stars are more numerous than gravitational microlensing events, but microlensing events have several signatures which are distinct from all known types of variable star: the brightenings have a precisely known symmetric shape described by only three parameters, they are achromatic, they will not repeat in a given star, and they should be distributed randomly among all stellar types and luminosities.

Griest *et al.*(1991) and Paczyński (1991) showed that the Galactic bulge also provides a promising target for microlensing searches. In particular, sources in the bulge provide sensitivity to low mass stars in the Galactic disk, and 'disk dark matter' in addition to a possible signal from halo dark matter. These contributions may potentially be separated by observations of fields at different galactic latitude and longitude, and also by comparing the mean timescale of the events.

Three groups are undertaking observational searches for microlensing, and all have reported detections of candidate events. Our MACHO collaboration has reported one striking candidate event in the LMC (Alcock *et al.* 1993); the EROS collaboration simultaneously reported two events in the LMC (Aubourg *et al.* 1993); and the OGLE collaboration has reported a total of six events in the Galactic bulge (Udalski *et al.* 1993, 1994a).

By good fortune, the galactic bulge is located $\sim 12^h$ away from the LMC in right ascension, hence it is well placed for observing when the LMC is at low altitude. Our group is presently the only one observing the LMC, the SMC and the Bulge at regular intervals. In this *Letter*, we report on a preliminary analysis of one of our bulge fields, containing $\approx 430,000$ stars with 130 observations from 1993 February to September, which has yielded four candidate microlensing events. In § 2, we provide an outline of the observations and photometric reduction; we summarize the selection and characteristics of the events in § 3, in § 4 we discuss microlensing detection efficiencies, and in § 5 we provide a brief discussion of some implications.

## 2. Observations and Data Analysis

The MACHO project has full-time use of the 1.27-meter telescope at Mount Stromlo Observatory, Australia. A system of corrective optics has been installed at the prime focus, giving a focal reduction to $f/3.9$ with a $1^o$ diameter field of view. A dichroic beamsplitter and filters provide simultaneous images in two passbands, a 'red' band (approx. 6300–7600 Å) and a 'blue' band (approx. 4500–6300 Å). Two very large CCD cameras are employed at the two foci; each contains a $2 \times 2$ mosaic of $2048 \times 2048$ pixel Loral CCD imagers. The pixel size is 15 $\mu m$ which corresponds to $0.63''$ on the sky, giving a sky coverage of 0.5 square degrees. Each chip has two read-out amplifiers, and the images are read out through a 16-channel system and written into dual-ported memory in the data acquisition computer. The readout time is 70 seconds per image, and the noise is $\sim 10$ electrons rms, with a gain of $\sim 1.9\, e^-/\mathrm{ADU}$. Images are saved on disk and Exabyte tape and a photometry code is automatically run. Details of the camera system are given by Stubbs *et al.* (1993).

The default exposure times are 300 seconds for LMC and SMC frames and 150 seconds for bulge frames. As of 1994 April, over 17,000 images have been taken with the system, of which about 11,000 are LMC, 1,000 SMC and 5,000 bulge. The frames are taken at standard sky positions, of which we have defined 82 in the LMC, 21 in the SMC and 75 in the bulge.

Photometric measurements from these images are made with a special-purpose code known as SoDoPHOT (Bennett *et al.* 1994), derived from DoPHOT (Schechter *et al.* 1993). First, one image of each field with good seeing and dark sky is chosen as a 'template image'. This is processed in a manner similar to a standard DoPHOT reduction except that after one color of the image has been reduced, the coordinates of the stars found in the first color are used as starting points for the positions of stars in the second color, which improves the star matching between colors. (The final positions of the matched stars are forced to be at a common position, after allowing for differential refraction.) This procedure provides a 'template' catalog of stellar positions and magnitudes for each field.

All other images are processed in 'routine' mode, which proceeds as follows. First the image is divided into 120 'chunks' of $\sim 512 \times 512$ pixels, and for each chunk $\sim 30$ bright stars are located and matched with the template. These stars are used to determine an analytic fit to the point spread function, a coordinate transformation and a photometric zero point. Then, all the template stars are subtracted from the image using the model PSF and coordinate transformation. Next, photometric fitting is carried out for each star in descending order of brightness, by adding the analytic model of the star back to the subtracted frame, and fitting with the model PSF and position. When a star is found to vary significantly from its template magnitude, it and its neighbors undergo a second iteration of fitting. For each star, the estimated magnitude and error are determined, along with 6 other parameters measuring the object 'type', the $\chi^2$ of the PSF fit, the crowding, the weighted fractions of flux removed due to bad pixels and cosmic rays, and the fitted sky value. The photometric error estimate is the formal PSF fit error (as in DoPHOT) with a 1.4% systematic error added in quadrature. This takes approximately 1 hour per image on a Sparc-10 for a field with 500,000 stars in each color. The set of photometric reductions for each field are re-arranged into a time-series for each star, and passed to an automated analysis to search for variable stars and microlensing candidates.

The observations discussed here cover a single field (MACHO field number 108) in the bulge, centered at $\alpha = 18h\ 01'\ 20''$, $\delta = -28°\ 17'\ 39''$ (J2000), i.e. $l = 2.30°, b = -2.65°$ in galactic coordinates. The template reduction contains 630,000 stars. We have analyzed 135 images of this field, covering the 190 nights from 1993 Feb 26 to 1993 Sep 03; observations were obtained on 113 of these nights, of which 22 nights have image pairs taken a few hours apart. A color-magnitude diagram is shown in Figure 1. Of the stars in the template, 431,700 had at least seven simultaneous satisfactory 'red' and 'blue' measurements and were subjected to further analysis.

## 3. Event Detection

The microlensing search through the light curve database proceeds in three stages: first, the time-series are convolved with a set of filters of various durations in order to search for peaks of any kind. Any lightcurve with a significant peak is tagged as a level 1 trigger. For these level 1 lightcurves a 5-parameter fit to a microlensing event is made, where the parameters are the unamplified red and blue fluxes, the peak amplification $A_{max}$, the time of maximum amplification $t_{max}$ and the event timescale $\hat{t}$. We define $\hat{t} = 2R_e/v_\perp$, where $R_e$ is the radius of the Einstein ring, and $v_\perp$ is the transverse velocity of the lens relative to the line-of-sight. Then, a set of statistics describing significance level, goodness of fit, achromaticity, crowding, temporal coverage of the event, etc. are calculated. Events above a modest significance level are tagged as level 1.5 events and are output as ASCII files, along with their associated statistics, and are then subjected to more rigorous selection criteria to search for microlensing candidates.

The analysis described here is somewhat different from the analyses we have performed on the LMC fields (Alcock et al, 1993) and should be regarded as preliminary. The LMC and bulge fields differ in many ways, but the most noticeable difference between this bulge field and the densest LMC fields was that artifacts due to systematic errors in the photometry occurred with a higher statistical significance, relative to the uncertainties assigned by the photometry. The detected star density in both of these fields is limited by crowding, but in the bulge fields we reach the crowding limit with stars that appear twice as bright as in the LMC bar fields. Thus, crowding-related systematic photometry errors can be expected to appear in brighter stars with a higher ratio of systematic error to noise in the bulge than in the LMC.

The majority of the level 1.5 trigger lightcurves contain small bumps attributable to these systematic errors. These are lightcurves of stars that can barely be distinguished from their nearest neighbors even in the best seeing images. Most of them are removed by cuts on the photometry output "crowding" parameters which measure the blending between nearby stellar images. Most of the of stars which have $V - R > 1.6$ are variables, and many are observed to have light curves which resemble low amplitude microlensing light curves. (See Fig. 3 for an example.) Unlike microlensing events, they generally do not remain at a constant brightness before and after the peak, but they can mimic microlensing over a limited duration. We therefore apply a cut $V - R < 1.6$ to our data in order to remove this potential source of background, eliminating 0.5% of our stars. Additional cuts were made on the microlensing curve fit $\chi^2$ (dof) < 3, the $\chi_p^2 < 4$ in the region of the fit peak, and the event timescale $\hat{t} < 200$ days. We also demanded that the the 'half maximum' points of the best fit lightcurve fall within our sample period. These cuts reduce the set of candidates to 5335 stars. The final two cuts on these 5335 stars are $A_{max} > 1.5$ and the 'normalised' $\Delta\chi_n^2 > 400$, defined as

$$\Delta\chi_n^2 = \frac{\chi_C^2 - \chi_{ML}^2}{\chi_{ML}^2 / N},$$

where $\chi_C^2$ is the (unreduced) $\chi^2$ for a 2-parameter fit to constant flux in each colour, $\chi_{ML}^2$ is the same for the 5-parameter fit to a microlensing event, and $N$ is the number of data points.

These cuts are shown in Figure 2. Four stars clearly stand out in this Figure, and these are found to be well fitted by microlensing lightcurves as shown in Figures 3 and 4. (Lightcurve files are available from alcock@sunlight.llnl.gov.) These form our sample of candidate 'events', and their characteristics are summarized in Table 1. It is encouraging that their variations show no evidence for color change, even though an explicit cut on chromaticity was not applied. (See Figure 4.) Plate 1 is a mosaic of images of the candidate events both unamplified and at peak amplification.

The light curves presented in the Figures are consistent with our expectations for gravitational microlensing. There are more points in the red data for events 1 and 4 because these source images fell on or near defects in our blue CCDs. We have cataloged these defects and when the photometry of an image is significantly affected (the primary effect being a loss of accurate error estimation), this photometric measurement is not used in searching for microlensing, nor in fitting a microlensing curve. The source image for event 4 was often near one such defect which may account for the apparent excess scatter in its blue data.

We conclude that the four events discussed above are *consistent* with microlensing. Follow-up spectral analysis and continued monitoring of these stars is underway. When we have analyzed more fields, found proper efficiencies, and developed a uniform set of selection criteria we should be able to make powerful statements about the nature of the Machos in the bulge and LMC lines-of-sight. However, even at this stage, some preliminary conclusions can be drawn, as will be discussed in § 5.

## 4. Detection Efficiency

In order to draw conclusions about the population of objects responsible for microlensing events, we must understand our microlensing detection efficiency. There are a number of factors that can prevent us from detecting microlensing events. The cuts described above will miss short timescale events that peak during periods of poor sampling, and they explicitly exclude long timescale events with $\hat{t} > 200$ days. In addition, events in the light curves of faint stars can fail to pass the $\Delta\chi^2$ cut if they are of fairly low amplitude or happen to peak during poor seeing, and some stars are explicitly removed by our crowding and color cuts. These "sampling" efficiencies can be measured by adding artificial microlensing events to stars in our light curve database and then checking how many of these are detected by our analysis.

We have run Monte Carlo simulations to determine our sampling efficiencies by adding fake microlensing light curves to 1% of the light curves in our database. For each simulation, we distribute "stars" of a unique mass M in a double exponential disk with scale height 325 pc and scale length 3.5 kpc.

| Event | RA (2000) | Dec (2000) | V | V-R | $t_{\max}$/days | $\hat{t}$/days | $A_{\max}$ | $u_{\min}$ | $\chi^2$ |
|---|---|---|---|---|---|---|---|---|---|
| 1 | 18 02 09.8 | -28 26 04 | 19.0 | 0.9 | 572.801(2) | 20.82(13) | 17.39(8) | 0.058 | 0.87 |
| 2 | 18 00 01.2 | -28 27 41 | 19.7 | 1.1 | 580.52(17) | 42.4(1.4) | 2.92(3) | 0.36 | 0.65 |
| 3 | 18 00 25.9 | -28 02 35 | 18.8 | 1.2 | 583.32(24) | 49.2(1.5) | 2.20(2) | 0.50 | 2.03* |
| 4 | 18 00 11.5 | -28 14 59 | 17.4 | 0.9 | 510.33(47) | 48.6(2.5) | 1.76(2) | 0.66 | 2.54 |

**Table 1:** Parameters of the events. Columns 2 & 3 show coordinates. Columns 4 & 5 show approximate magnitude and color of the lensed stars, ( using an approximate transformation from our non-standard passbands ). Columns 6-9 show the parameters of the best-fit microlensing models: time of peak amplification (days from 1992 Jan 02), the event duration, the peak amplification factor, and the impact parameter, with the *formal* one sigma errors (derived from the covariance matrix of the fit) in the last one or two significant digits shown in parantheses (we believe these are underestimates of the true errors). Column 10 is the $\chi^2$ per degree of freedom for the microlensing fit. * The $\chi^2$ for event 3 drops to 1.02 if one set of measurements at day 452.7 is dropped.

For masses in the $0.1 M_\odot$ to $1.0\ M_\odot$ range, we recover at most 40% of the fake events with input peak amplifications $> 1.34$ using the cuts described above. For lens masses $< 0.1 M_\odot$ or $\gg 1 M_\odot$, the efficiency is substantially less than this.

Equally important but more difficult to measure is the "photometric" efficiency. The photometric efficiency refers to the effect of photometric errors which are not decribed by the reported error estimates, and the fact that many of the "stars" that we have identified are actually blends of 2 or more stars which have separations $\lesssim 1"$. These efficiencies can be measured by adding artificial stars with artificial microlensing events to the images and measuring what fraction of the artificial events can be recovered. The determination of our photometric efficiency is currently in progress, and will be reported on in a subsequent paper. The efficiency results reported in this paper do not included these results and so will necessarily be only upper limits.

We emphasize that both the sampling and blend efficiencies are likely to differ substantially for our observations of the bulge and the LMC due to significant differences in the luminosity functions and in our time sampling of the different fields.

## 5. Discussion

As discussed above, and summarized in Table 1, we have found four candidate microlensing events toward the galactic bulge, with durations between 20 and 50 days, after monitoring $\sim 430,000$ stars over 187 days.

It is of course of great interest to make a comparison with theoretical predictions of microlensing rates. A proper estimate of the number of events we expect to see and the interpretation of these depend critically upon an accurate knowledge of our detection efficiencies. However, we can bracket the expected rate by using the variable $\epsilon$ to parameterize our ignorance of the efficiencies. We can compare our event rate with predictions from previously studied populations distributed in the halo, disk (Griest et al., 1991, Paczyński 1991), and a massive spheroid (Giudice et al., 1993).

For the one bulge field our "total exposure" $E = (431,708 \text{stars})(186.8 \text{days}) = 8.06 \times 10^7$ star-days $= 2.21 \times 10^5$ star-years. Then the expected number of events is $N_{exp} = \epsilon E \Gamma$, where $\Gamma$ is the event rate predicted by a theoretical model of the Macho positions and velocities. By convention, the event rate refers to the rate for all events with a peak magnification greater than 1.34, and the effect of a threshold $A_{\max} \neq 1.34$ is folded into the efficiency estimate.

For lensing of bulge stars by disk stars with a Scalo (1986) mass function, Griest et al. (1991) predicted an expected number of events

$$N_{exp}(\text{disk stars}) = \epsilon r_\eta (2.2 - 7.4 \times 10^{-6})(2.21 \times 10^5) = \epsilon r_\eta (0.5 - 1.7) = \epsilon(0.8 - 2.6)$$

events, where $r_\eta \sim 1.6$ is an approximate scaling from $b = -4°$ used by Griest et al. to $b = -2.65°$ where our field is located, and where the spread in predicted rate comes from uncertainties in the mass function of faint disk stars. Thus with an efficiency $\epsilon < 0.4$, we have $N_{exp}(\text{disk stars}) < 1.1$ events. With four events, the 95% confidence level lower limit on the mean number of events for the total exposure reported here is 1.37 (0.82 if only three events are included). Thus, our rate is significantly higher than the prediction of Griest et al. if all 4 events are due to microlensing of bulge stars by disk stars.

To estimate the total optical depth that we've observed, we could sum up the amount of time we detect microlensing with an amplification greater than 1.34. The optical depth is then the ratio of this time to E, with $\epsilon$ appropriately taken into account. Alternatively, we can take each detected event and sum the amount of time that the average event with the same $\hat{t}$ would spend with A $\geq 1.34$. This gives

$$\tau = \frac{\pi}{4} \frac{1}{\epsilon E} \sum_{i=1,4} \hat{t}_i$$
$$= 1.57 \times 10^{-6}/\epsilon \gtrsim 3.9 \times 10^{-6}$$

where $\hat{t}_i$ refers to $\hat{t}$ for the $i$th event. We have evaluated the statistical uncertainty in this number by drawing random events from our Monte Carlo simulations using varying masses for the lensing objects and demanding that at least 5% of the random samples have at least 4 events *and* $\sum_{i=1,4} \hat{t}_i \geq 161$ days. Demanding that the simulated events satisfy both the 161 day and 4 event criteria at the 95% confidence level yields a limit of $\tau > 1.3 \times 10^{-6}$. (The top 3 events alone imply $\tau > 8 \times 10^{-7}$.)

We have computed the optical depth for double exponential disk models with a wide range of parameters, and find a good correlation between the optical depth and $v_d$, the local circular velocity due to the disk. For any plausible parameter choices, we find $\tau \lesssim 10^{-6}(v_d/140\text{kms}^{-1})^2$. The Griest *et al.* optical depth estimates range from $\tau = 5 \times 10^{-7}$ for a minimal 'stellar' disk to $\tau = 3 \times 10^{-6}$ for the maximal disk (including a 'dark' component). Other populations may also contribute to the optical depth toward the bulge: The galactic halo is expected to contribute only $\tau \sim 10^{-7}$ (Griest 1991) unless the core radius is quite small. A massive spheroid might contribute $\tau \approx 6 \times 10^{-7}$ (Giudice *et al.*, 1993), and we estimate the contribution of bulge stars microlensing other bulge stars (using the model of Bahcall and Soneira 1980) to be $\tau \approx 5 \times 10^{-7}$. (Kiraga and Paczyński (1994) find a similar result.) Finally, we should point out that a proper calculation of the expected optical depth should include information about the distribution of the source stars along the line of sight. Perhaps 10-20% of the stars we observe might be in the foreground where microlensing would be much less likely while an unknown fraction of the stars we see may actually lie behind the bulge which could substantially increase the microlensing probability.

Of the models considered, only the maximal disk models with $v_d > 200$km/s can come close to our best fit value of $\tau \gtrsim 4 \times 10^{-6}$, but due to the small number of events, the 'minimal disk only' model is the only model which can *formally* be ruled out at the 95% confidence level.

Using the formula from Griest *et al.*, it is also possible to find probable masses for the candidate events once an assumption is made about the population to which they belong. For a disk population, $<m> = \hat{t}^2/(85\text{days})^2 M_\odot$, so $<m> \approx 0.06 M_\odot$ for $\hat{t} = 20.8$ days, and $<m> \approx 0.23 M_\odot$ for $\hat{t} = 40$ days (the average of all 4 events). The inferred most likely masses of the three lower amplification events (Events 2–4) are consistent with faint disk stars (or also with a brown dwarf population of disk dark matter). The inferred most likely mass for Event 1 is somewhat lower, but these mass results are of dubious significance given the small number statistics and other model uncertainties.

In summary, we have detected four candidate microlensing events by monitoring 430,000 stars in the Galactic bulge for 190 days. If all of the events are due to microlensing, the event rate 1) appears to exceed the predictions of the 'minimal disk' model of Griest *et al.* (1991), and 2) is probably consistent with 'maximal disk' models and with the results from Udalski *et al.* (1994). This may have important implications for models of Galactic structure and indirectly for predictions of microlensing toward the Magellanic Clouds by a halo dominated by MACHOs. Clearly, improved statistics are required, as well as observations covering a range of galactic coordinates. At present, we have more than an order of magnitude more bulge data awaiting analysis, and we plan to present more detailed results in a future publication.

We are very grateful for the skilled support given our project by the technical staff at the Mt. Stromlo Observatory. Work performed at LLNL is supported by the DOE under contract W7405-ENG-48. Work performed by the Center for Particle Astrophysics on the UC campuses is supported in part by the Office of Science and Technology Centers of NSF under cooperative agreement AST-8809616. Work performed at MSSSO is supported by the Bilateral Science and Technology Program of the Australian Department of Industry, Technology and Regional Development. KG acknowledges a DOE OJI grant, and CWS and KG thank the Sloan Foundation for their support.

# Figure Captions

Figure 1    Color magnitude diagrams of the regions surrounding our four events. The location of the source star for each event is indicated by a large dot. Each panel shows the color magnitude diagram for a 5.3 x 5.3 arcmin region containing the appropriate source star: (a) Event 1, (b) Event 2, (c) Event 3, and (d) Event 4.

Figure 2    This is a scatter plot of the $\Delta\chi^2$ vs. best fit $A_{\max}$ for the 5335 stars passing our other cuts. The four microlensing candidates fall above the cuts in the upper right region of the diagram.

Figure 3    The full light curves are plotted for events 1-4 and a red variable with $V - R = 1.8$ (background). Roughly ten stars with $V - R > 1.6$ pass the $\Delta\chi^2$ cut, but all have $A_{\max} < 1.5$. Time is in days from 1992 Jan 2 00:00 UT, and the amplitude is given in linear units normalized to the median detected flux from the star.

Figure 4    'Close-ups' of the event light curves are plotted showing the best microlensing fit curves and the $A_{blue}/A_{red}$ ratio to test for chromaticity. The points marked with a cross for event 1, have been removed by the microlensing search software because a significant fraction of the PSF of the star is overlapped by bad columns. These points have not been used in the microlensing fit nor the $A_{blue}/A_{red}$ ratio.

Plate 1    A mosaic of an unamplified and an amplified image for each event. The bottom row shows the source stars in relatively good seeing at their mean magnitude. The top row shows the brightest point sampled during the event. Each box is approximately 28 x 28 arcsec; north is to the left and east is down.

# References

Alcock, C., Akerlof, C.W., Allsman, R.A., Axelrod, T.S., Bennett, D.P., Chan, S., Cook, K.H., Freeman, K.C., Griest, K., Marshall, S.L., Park, H.-S., Perlmutter, S., Peterson, B.A., Pratt, M.R., Quinn, P.J., Rodgers, A.W., Stubbs, C.W., & Sutherland, W., 1993, *Nature*, **365**, 621.

Aubourg, E., Bareyre, P., Brehin, S., Gros, M., Lachieze-Rey, M., Laurent, B., Lesquoy, E., Magneville, C., Milsztajn, A., Moscosco, L., Queinnec, F., Rich, J., Spiro, M., Vigroux, L., Zylberajch, S., Ansari, R., Cavalier, F., Moniez, M., Beaulieu, J.-P., Ferlet, R., Grison, Ph., Vidal-Madjar, A., Guibert, J., Moreau, O., Tajahmady, F., Maurice, E., Prevot, L., & Gry, C., 1993, *Nature*, **365**, 623

Bahcall, J.N. & Soneira, R.M., 1980, *Ap. J. Supp.*, **44**, 73.

Bennett, D. P. *et al.*, 1994, in preparation

Giudice, G.F., Mollerach, S., & Roulet, E., 1993, (preprint: CERN-TH.7127/93)

Griest, K., 1991, *ApJ*, **366**, 412

Griest, K., *et al.* 1991, *ApJ Lett.*, **372**, L79

Kiraga M., and Paczyński, B. 1994, *ApJ Lett.*, in press.

Paczyński, B, 1986, *ApJ*, **304**, 1

Paczyński, B. 1991, *ApJ Lett.*, **371**, L63

Scalo, J.M., *Fund. Cosmic Phys.*, **11**, 1

Stubbs, C., Marshall, S.L., Cook, K.H., Hills, R., Noonan, J., Akerloff, C.W., Axelrod, T.S., Bennett, D.P., Dagley, K., Freeman, K.C., Griest, K., Park, H.-S., Perlmutter, S., Peterson, B.A., Quinn, P.J., Rodgers, A.W., Sosin, C., & Sutherland, W., 1993, *SPIE Proceedings*, **1900**, 192

Schechter, P.L., Mateo, M., & Saha, A., 1994, *PASP*, **105**, 1342

Udalski, A., Szymanski, M., Kaluzny, J., Kubiak, M., Krzeminski, W., Mateo, M., Preston, G.W., & Paczynski, B., 1993, *Acta Astronomica*, **43**, 289

Udalski, A., *et al.*, 1994, *ApJ*, **426**, L69.